\documentclass[
	superscriptaddress, 12pt,amsmath,amssymb, preprintnumbers
	]{revtex4-1}

\newcommand{\figurewidth}{.65\columnwidth}
\usepackage{graphicx}
\usepackage{siunitx}	
\usepackage{amssymb}
\usepackage{datetime}
\usepackage{nth}
\usepackage{pgffor}
\usepackage[utf8]{inputenc}
\usepackage{pdfpages}

\makeatletter
\AtBeginDocument{\let\LS@rot\@undefined}
\makeatother

\newcommand{\TiSe}{TiSe$_2$}
\newcommand{\NbSe}{NbSe$_2$}
\newcommand{\VSe}{VSe$_2$}

\newcommand{\TCDW}{\ensuremath{T_{\text{CDW}}}}

\newcommand{\RH}{\ensuremath{R_{\text H}}}

\newcommand{\rhoxx}{\ensuremath{\rho_{xx}}}
\newcommand{\rhoxy}{\ensuremath{\rho_{xy}}}
\newcommand{\MR}{\ensuremath{\text{MR}}}
\newcommand{\dd}{\text{d}}										

\newcommand{\kB}{\ensuremath{k_{\text B}}}
\newcommand{\muh}{\ensuremath{\mu_{\text h}}}
\newcommand{\mue}{\ensuremath{\mu_{\text e}}}
\newcommand{\nh}{\ensuremath{n_{\text h}}}
\newcommand{\nel}{\ensuremath{n_{\text e}}}

\newcommand{\mel}{\ensuremath{m_{\text e}^{\star}}}
\newcommand{\me}{\ensuremath{m_{\text e}}}

\newcommand{\rhoO}{\ensuremath{\rho_{0}}}
\newcommand{\K}{\ensuremath{K}}

\newcommand{\DeltaCDW}{\ensuremath{\Delta_{\text{CDW}}}}
\newcommand{\sigmah}{\ensuremath{\sigma_{\text h}}}
\newcommand{\sigmae}{\ensuremath{\sigma_{\text e}}}
\newcommand{\sigmat}{\ensuremath{\sigma_{\text t}}}
\newcommand{\rhot}{\ensuremath{\rho_{\text t}}}

\begin{document}

%
%


\title{Fermi surface reconstruction and electron dynamics at the charge-density-wave transition in \TiSe}

\author{Patrick Knowles}
\affiliation{HH Wills Laboratory, University of Bristol, Bristol, BS8 1TL, UK}

\author{Bo Yang}
\affiliation{HH Wills Laboratory, University of Bristol, Bristol, BS8 1TL, UK}

\author{Takaki Muramatsu}
\affiliation{HH Wills Laboratory, University of Bristol, Bristol, BS8 1TL, UK}

\author{Owen Moulding}
\affiliation{HH Wills Laboratory, University of Bristol, Bristol, BS8 1TL, UK}

\author{Jonathan Buhot}
\affiliation{High Field Magnet Laboratory, University of Radboud, Nijmegen, NL}

\author{Charles Sayers}
\affiliation{Centre for Nanoscience and Nanotechnology, Department of Physics, University of Bath, BA2 7AY Bath, UK}

\author{Enrico Da Como}
\affiliation{Centre for Nanoscience and Nanotechnology, Department of Physics, University of Bath, BA2 7AY Bath, UK}

\author{Sven Friedemann}
\email{Sven.Friedemann@bristol.ac.uk}
\affiliation{HH Wills Laboratory, University of Bristol, Bristol, BS8 1TL, UK}

\date{\today}


\keywords{\TiSe, charge-density-wave, quantum critical point, Fermi surface, magnetoresistance}

\begin{abstract}
The evolution of the charge carrier concentrations and mobilities are examined across the charge-density-wave (CDW) transition in \TiSe. Combined quantum oscillation and magnetotransport measurements show that a small electron pocket dominates the electronic properties at low temperatures while an electron and hole pocket contribute at room temperature. At the CDW transition, an abrupt Fermi surface reconstruction and a minimum in the electron and hole mobilities are extracted from two-band and Kohler analysis of magnetotransport measurements.  The minimum in the mobilities is associated with the overseen role of scattering from the softening CDW mode. 
With the carrier concentrations and dynamics dominated by the CDW and the associated bosonic mode, our results highlight \TiSe\ as a prototypical system to study the Fermi surface reconstruction at a density-wave transition.
\end{abstract}

\maketitle

%
%
The electronic properties of transition metal dichalcogenides (TMDs) are of fundamental and practical interest. Many TMDs can be tuned between semimetallic, semiconducting, and insulating behaviour and thus allow to access a plethora of different electronic characteristics. In addition, ordered states, e.g.\ due to charge-density-wave (CDW) formation \cite{Wilson1975a,Wilson1969} or superconductivity \cite{Morosan2006a,Kusmartseva2009} are present in many members of the family with open questions on the underlying mechanism. 
Many of these TMDs can be exfoliated to atomic monolayers providing new tuning parameters and novel physics through the reduced dimensionality \cite{Xi2015,Xi2015a, Singh2017}.

\TiSe\ is a prototypical material for strong electronic interactions driving the CDW formation via a condensation of excitons, i.e. pairs of electrons and holes \cite{DiSalvo1976,Wilson1978}. Experimental and theoretical work have confirmed the relevance of the excitonic mechanism \cite{Hellmann2012,Cercellier2007,Kogar2017} which is widely accepted to work in cooperation with strong electron-phonon coupling \cite{Hedayat2019,Porer2014,Wezel2010a}. 

Above the CDW transition temperature $\TCDW = \SI{202}{\kelvin}$, \TiSe\ is characterised by small carrier concentrations stemming from up to three selenium-derived hole-like bands with cylindrical topology at the $\Gamma$-point and a titanium-derived electron band with distorted and tilted ellipsoid topology present with 3-fold multiplicity at the L-point \cite{Bianco2015,Rasch2008,Watson2019,Pillo2000} \footnote{We use the high-temperature notation of the Brilluoin zone throughout the manuscript. In the low-temperature phase the high-temperature L-point folds back onto the high-temperature $\Gamma$-point}. Whether these bands overlap in energy or have a band gap remains uncertain. Either way, the overlap or gap is small or comparable to thermal energies down to \SI{50}{\kelvin}. 

Below the CDW transition temperature, the electronic structure of \TiSe\ is dominated by a small electron pocket as shown by ARPES measurements \cite{Watson2019}. 
Knowledge of how the electronic structure and electron dynamics evolve upon thermally melting the CDW in equilibrium, however, are outstanding. Several studies suggested that the Fermi surface reconstruction from the CDW order and scattering associated with the CDW mode has a negligible effect on the electronic structure and dynamics \cite{DiSalvo1976,Watson2019a,Pillo2000,Monney2010}. Rather a dominance of thermal occupation effects was suggested.
In the past, studies of the charge carrier concentration were based on a single-band analysis of Hall effect and optical reflectivity measurements despite the evidence for two bands being present while no measurements distinguishing the electron and hole dynamics across \TCDW\ have been reported \cite{Velebit2016,Li2007a}. 
Here, we use high-resolution magnetotransport and quantum oscillation (QO) measurements to extract the temperature dependence of the charge carrier concentration and mobility of the electron and hole band.
We directly observe one quasi-ellipsoid Fermi surface at low temperatures which is identified as an electron pocket. This electron pocket and a hole pocket grow rapidly above \SI{150}{\kelvin} showing evidence of an abrupt Fermi surface reconstruction and gapping of \SI{75}{\percent} of the charge carrier concentration in the CDW state. At the same time, we observe a minimum in the mobility on both the electron and hole pocket at \TCDW\ highlighting the importance of scattering from the CDW forming phononic and/or electronic modes.

Single crystals of \TiSe\ were grown by chemical vapour transport at \SI{580}{\celsius} as detailed in sec.~%
~SI
of the Supplemental Material, and show a CDW transition at $\TCDW=\SI{202}{\kelvin}$  consistent with other studies of high-quality samples\cite{Taguchi1981, Huang2017, Moya2019,DiSalvo1976,Hildebrand2014}.


%
%

\begin{figure}%
\includegraphics[width=\figurewidth]{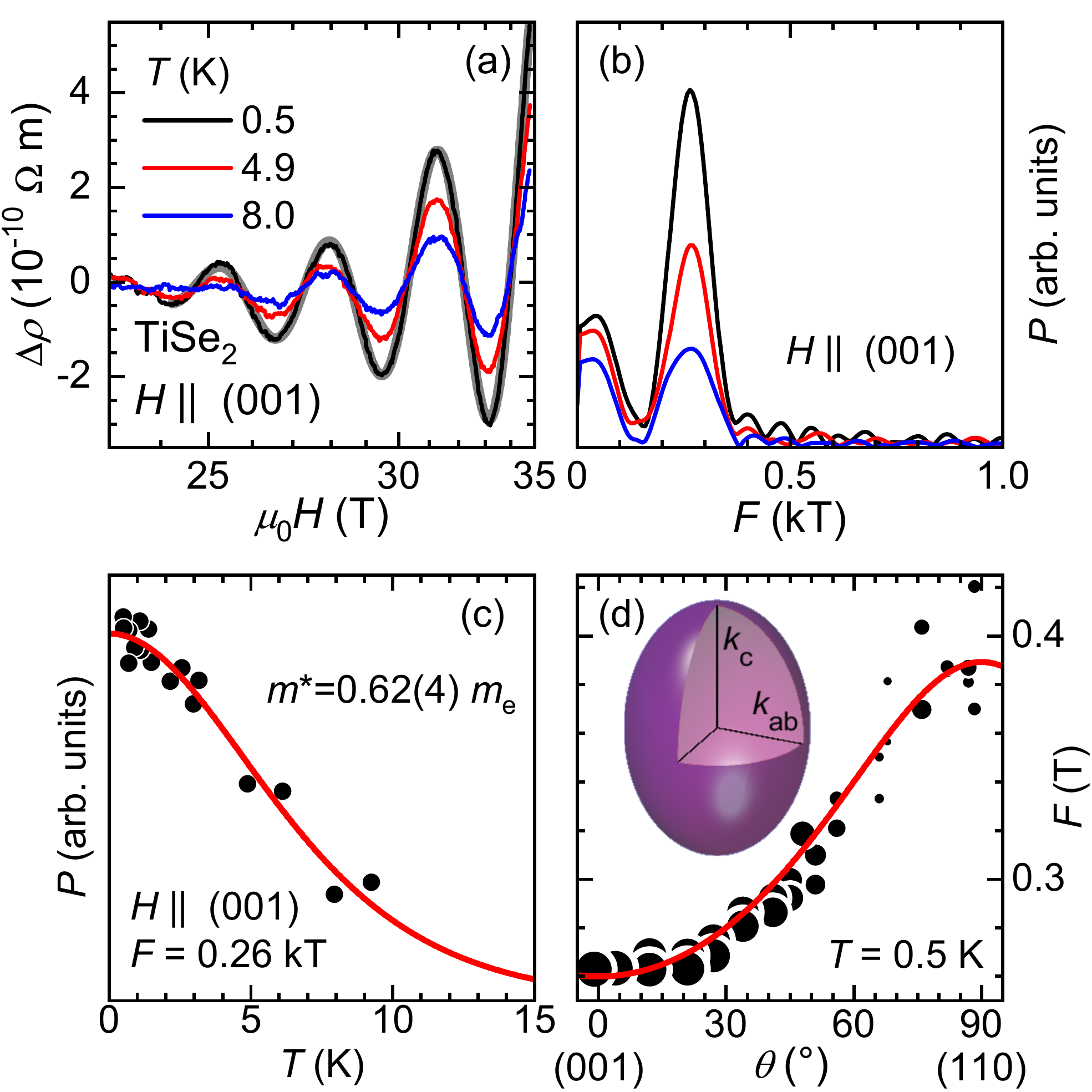}%
\caption{Quantum oscillations in \TiSe. (a) The oscillatory component of the resistivity $\Delta \rho$ obtained after subtraction of a linear background. The thick grey line is a fit of a single frequency with Dingle damping factor and quadratic background correction to the data at \SI{0.5}{\kelvin}. (b) Fourier transform spectrum. (c) Temperature dependence of the amplitude. The red solid line is a Lifshitz–Kosevich fit to the data with an effective mass of $\mel=\SI{0.62(4)}\,\me$, where \me\ denotes the bare electron mass. (d) Dependence of the oscillation frequency on zenith angle $\theta$ for fixed azimuthal angle $\phi=\SI{30}{\degree}$. Crystallographic directions are indicated on abscissa. Symbol size 
indicates oscillation amplitude. Solid red line represents a fit using an ellipsoid Fermi surface model as depicted in inset and described in 
section S~III of the Supplemental Material.}%
\label{fig:QO}%
\end{figure}

The low-temperature electronic structure of our \TiSe\ samples is dominated by an electron pocket as evident from the combination of quantum oscillation measurements and magnetotransport. The QO data shown in Fig.~\ref{fig:QO} reveal a single frequency $F=\SI{0.26}{\kilo\tesla}$ for magnetic fields parallel to the crystallographic $c$ direction, i.e.\ an orbit parallel to the basal plane. The increase of this frequency for orbits out of plane is well fitted by an ellipsoid shape 
(see Fig.~\ref{fig:QO}(d) and 
S~III
of the Supplemental Material). 
This pocket is naturally associated with the L-point electron pocket observed by ARPES studies \cite{Watson2019}. 


\begin{figure*}%
\includegraphics[width=\textwidth]{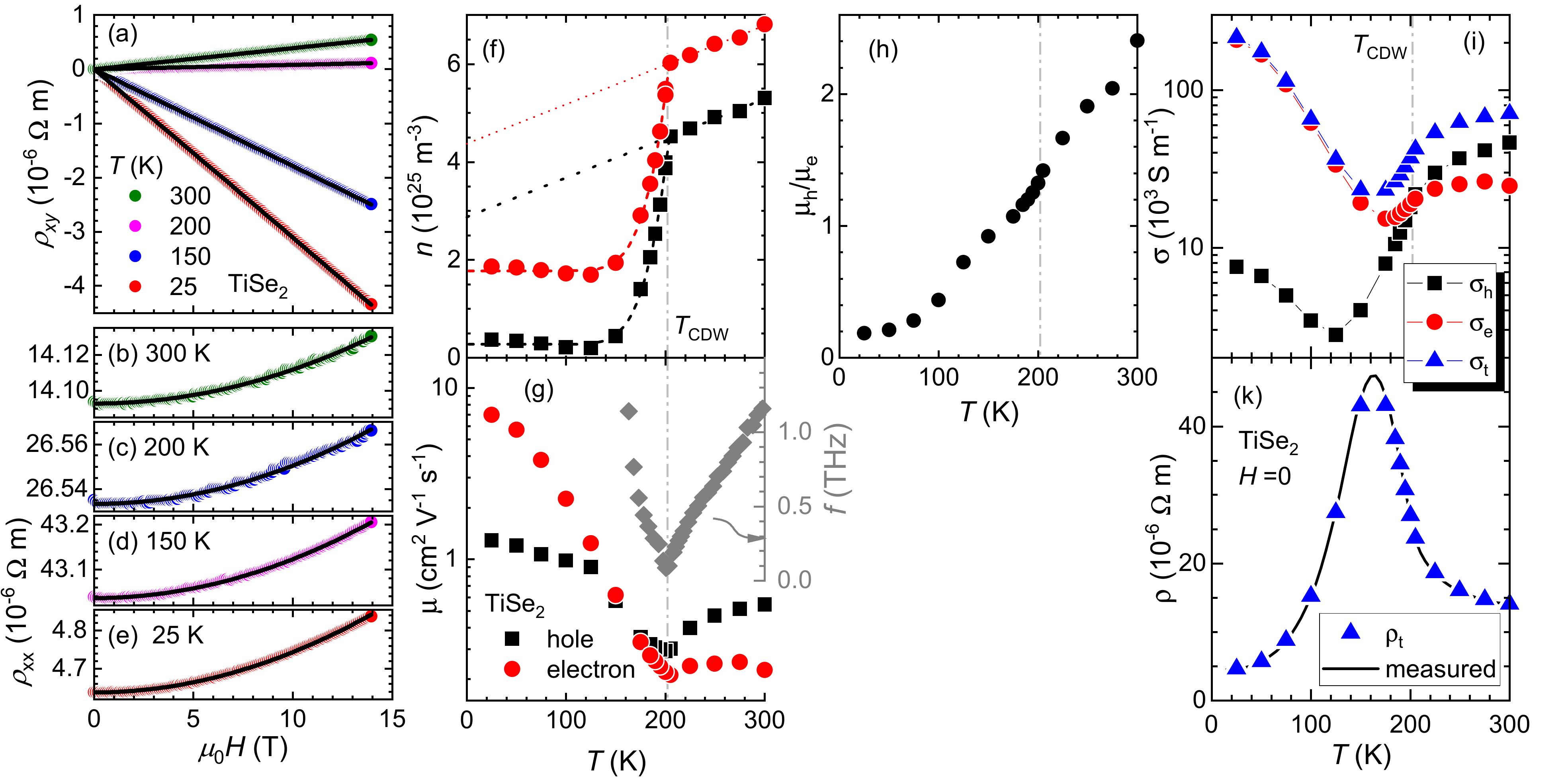}%
\caption{Two-band analysis of magnetotransport in \TiSe. Transversal (a) and longitudinal (b)-(e) resistivities of \TiSe\ at selected temperatures. Solid lines correspond to simultaneous two-band fits of both resistivities 
as described in 
S~II
of the Supplemental Material giving
charge carrier concentrations (f) and mobilities (g) of the hole and electron band. (g) includes frequency of $L$-mode phonon from Ref \cite{Holt2001} (adjusted for the reduced \TCDW\ of \cite{Holt2001}). 
Dotted lines in (f) represent linear fits to $\nel(T)$ and $\nh(T)$ for $T\geq\TCDW$. 
Dashed lines represent fits of the form $n_{\text{e,h}}(T)=n_{(\text{e,h})}(T=0)+A\exp(-2\Delta/\kB T)$.
(h) Ratio of hole and electron mobility. 
Vertical dash-dotted line indicates \TCDW. Standard errors as obtained from the non-linear least-squares fits are smaller than symbol size. 
(i) Separate zero-field electron \sigmae\ and hole band conductivity \sigmah\ have been calculated from the mobilities and charge carrier concentrations shown in (f) and (g). Total conductivity is calculated as $\sigma_{\text t} = \sigmae + \sigmah$. Note the logarithmic scale. (k) Measured zero-field resistivity (solid line) compared to the calculated total resistivity $\rho_{\text t} = 1/\sigma_{\text t}$ (triangles). Solid line the represents measured resistivity.
}%
\label{fig:MB}%
\end{figure*}

Magnetotransport measurements confirm the electron-dominated character of our samples at low temperature as can be seen from the negative Hall resistivity \rhoxy\ in Fig.\ref{fig:MB}(a). Consistent with earlier reports \cite{DiSalvo1976,Campbell2019}, \rhoxy\ changes sign smoothly around \SI{190}{\kelvin} 
(cf.\ Fig.~S2 of the Supplemental Material). 
The magnetoresistivity $\rhoxx(B)$ shown in Fig.~\ref{fig:MB}(b-e) is small and positive. 
We use a two-band model with one electron and one hole band to simultaneously fit $\rhoxx(\mu_0 H)$ and $\rhoxy(\mu_0 H)$ as described in section S~II of the Supplemental Material. The two-band fits shown as solid lines in Fig.~\ref{fig:MB} describe the data very well over the full temperature range.

The charge carrier concentrations \nel, \nh\ and mobilities \mue, \muh\ for the electron and hole band respectively are shown in Fig.~\ref{fig:MB} (f) and (g). The low-temperature value of \nel\ is in good agreement with our QO data and previous ARPES as well as heat capacity studies as summarised in Tab.~S1 of the supplementary information. 
Thus the association of the observed QO frequency with the electron pocket is confirmed. 
The electron pocket observed in our QO measurements accounts for virtually the full low-temperature electronic heat capacity (cf.\ 
Tab.~S1%
) highlighting the dominance of the electron pocket at low $T$.
The hole concentration extracted within the two-band model is very small at lowest temperatures and is likely to correspond to impurity states as indicated by the low hole mobility. 
We note that a free-electron single-band model cannot describe the low-temperature magnetotransport. Most notably, the Hall coefficient $\RH=\SI{-3.1e-7}{\meter\cubed\per\coulomb}$ measured at lowest temperatures does not match with a free electron  estimate for the electron pocket based on the QO results of $\RH^{\text {QO}} = \SI{-1.75e-7}{\meter\cubed\per\coulomb}$. 

At room temperature the electron and hole concentrations \nel\ and \nh\ are comparable in magnitude (cf.\ Fig.~\ref{fig:MB}(f)).
Above \TCDW, \nel\ and \nh\ are associated with the 3D-like electron pocket at the L point and the 2D-like hole pocket at the $\Gamma$ point as seen by ARPES studies \cite{Rasch2008,Watson2019}. The small linear temperature dependence of $\nel(T)$ and $\nh(T)$ above \SI{210}{\kelvin} (dotted lines in Fig.~\ref{fig:MB}(f)) is attributed to the varying thermal occupation of the two bands similar to the model of Watson et al. \cite{Watson2019a}. 
The slope of $\nel(T)$ and $\nh(T)$ at $T\geq\TCDW$ suggests a mass comparable to the free-electron mass for the electron band. The linear behaviour at high temperatures extrapolates to finite intercepts for both bands - these finite intercepts suggest a band overlap, i.e. not a gap, in the high-temperature phase above \TCDW.


The charge carrier concentrations show a sharp drop below \TCDW\ and saturate below \SI{150}{\kelvin}. The drop is  associated with condensation of electrons and holes into the CDW pair state and consequently with a Fermi surface reconstruction. 
The difference $\Delta n_{\text{e,h}} = n_{\text{e,h}}(\TCDW)-n_{\text{e,h}}(T=0) = \SI{4.2(2)e25}{\per\meter\cubed}$ marks the loss of charge carriers and thus the density of electron-hole pairs. 
For $T<\TCDW$, $\nel(T)$ and $\nh(T)$ can be described by activated behaviour as shown by  dashed lines in Fig.~\ref{fig:MB}(f).Fits of  exponential form yield a gap $\DeltaCDW = \SI{75(1)}{\milli\electronvolt}$, an energy scale consistent with $\TCDW = \SI{202}{\kelvin}$. The fact, that the exponential form fits the data even close to \TCDW\ suggests a finite gap up to \TCDW. A finite gap at \TCDW\ has indeed been seen in ARPES studies \cite{Chen2016b, Mottas2019, Monney2010b} where the total gap $\Delta =\Delta_{\text{CDW}} + \Delta_{\text{off}}$ is a sum of a BCS-like temperature dependent gap $\Delta_{\text{CDW}}$ with an onset at \TCDW\ on top of a weakly temperature dependent offset $\Delta_{\text{off}} \approx \SI{70}{\meV}$.
Our exponential fits are dominated by the temperature dependence of $n(T)$ just below \TCDW\ where $\Delta \approx \Delta_{\text{off}}$. The exponential form of $\nel(T)$ and $\nh(T)$ up to \TCDW\ and the good agreement of the gap value with $\Delta_{\text{off}}$ suggests a finite gap present above \TCDW\ potentially due to fluctuating electron-hole pairs that condense at \TCDW\ consistent with ARPES studies finding small intensity from backfolded bands above \TCDW\ \cite{Cercellier2007}.

\begin{figure}%
\includegraphics[width=\figurewidth]{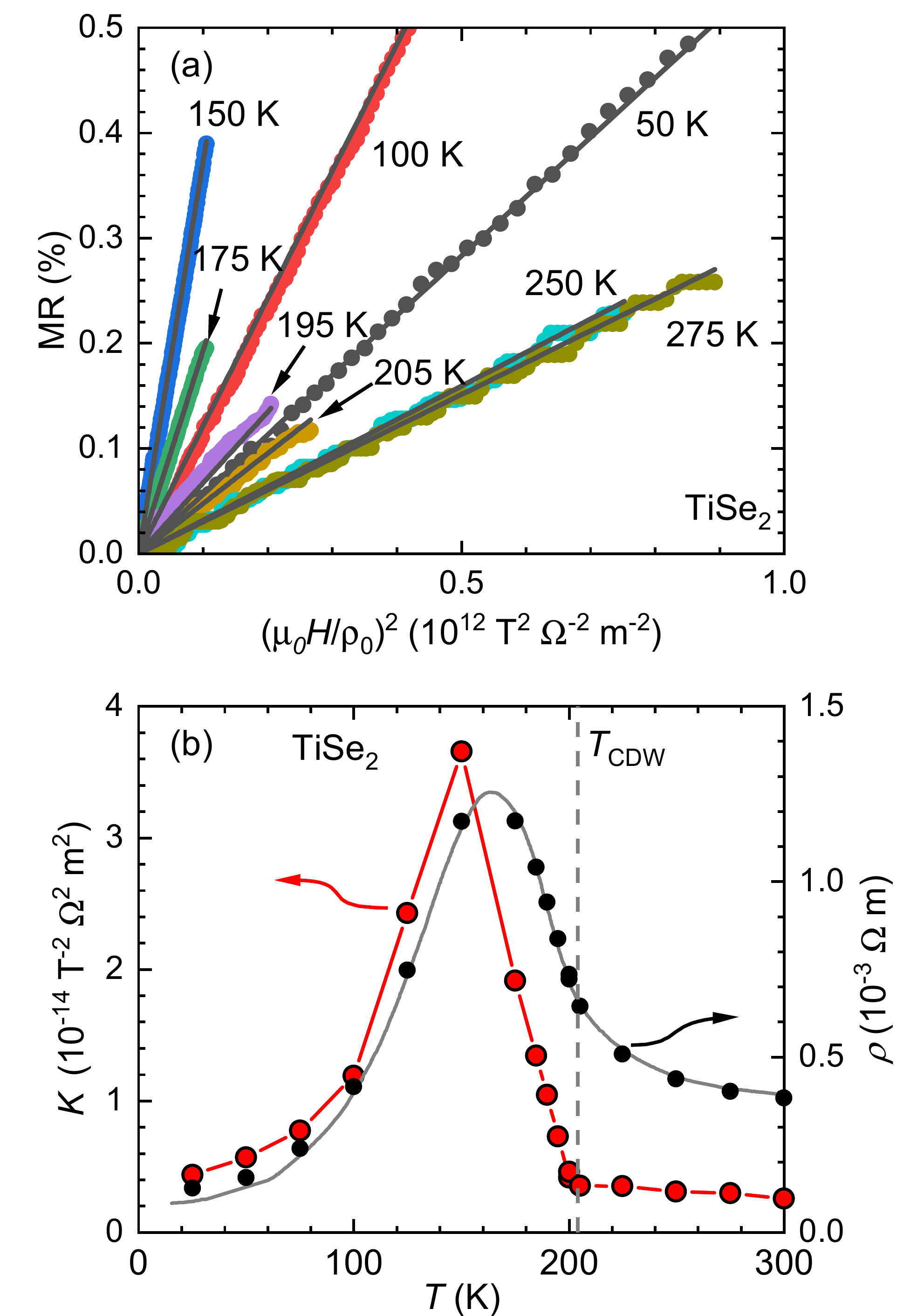}%
\caption{Kohler analysis of the magnetoresistance in \TiSe. 
(a) Magnetoresistance $\MR = (\rho-\rhoO)/\rhoO$ plotted against $(\mu_0 H / \rhoO)^2$ where $\rhoO=\rho(T,H=0)$ 
Solid lines are linear fits of $\MR = \K\ (\mu_0 H)^2$. 
(b) The Kohler slope $\K(T)$ (left axis) is shown with the solid red line as a guide to the eye. 
The zero-field resistivity $\rhoO(T)$ (black circles, right axis) is shown together with $\rho(T)$ from Ref.~\cite{DiSalvo1976} (solid line) where the latter has been scaled to our data.
}%
\label{fig:Kohler}%
\end{figure}

The Fermi-surface reconstruction is further supported by the Kohler analysis \cite{Pippard1989} presented in Fig.~\ref{fig:Kohler}. The magnetoresistance follows a quadratic field dependence at $T\geq\SI{50}{\kelvin}$. However, the quadratic coefficient (Kohler slope) \K\ shows a pronounced temperature dependence (cf.\ Fig.~\ref{fig:Kohler}(b)). Above \TCDW, \K\ is virtually constant and accordingly curves of \MR\ vs $(\mu_0 H / \rhoO)^2$ collapse in Fig.~\ref{fig:Kohler}(a). Below \TCDW, however, \K\ rises very abruptly by more than an order of magnitude, passes a maximum at \SI{150}{\kelvin} and saturates at a low-temperature value about double the room temperature value. 

Kohler scaling and violations thereof have been observed in other CDW systems: In VSe$_2$ and \NbSe, separate Kohler scaling is present below and above \TCDW\ with a small difference in slope at \TCDW\ \cite{Xue2020,Noto1980}. In Ta$_2$NiSe$_7$ and NbSe$_3$, Kohler scaling is only obeyed above \TCDW. In underdoped cuprate superconductors, Kohler scaling is observed at low temperature throughout \TCDW\ \cite{Chan2014}. Our results on \TiSe\ show a larger change in $K$ compared to other compounds because a larger fraction of the Fermi surface is affected by the CDW.

Our data suggest the Fermi surface reconstruction is the main reason for the violation of Kohler scaling in \TiSe.
The sharp rise and strong temperature dependence of \K\ below \TCDW\ can be due to (i) a reconstruction of the Fermi surface, (ii) an abrupt change in the anisotropy of the scattering time on the individual bands, or (iii) an abrupt change in the ratio of the scattering times of the electron and hole band, or a combination of the three. 
(iii) can be ruled out 
as this would manifest as an abrupt change of $\muh/\mue$  which is not observed (Fig.~\ref{fig:MB}(h)). 
(ii) may contribute to the change of \K\ but it is unlikely to be the primary cause. For the violation of Kohler's law to be dominated by changes to scattering time anisotropy, a drastic and abrupt change to the phonon spectrum would be required. This is unlikely to occur independent of the Fermi surface reconstruction. A moderate change of the phonon spectrum may occur as a consequence of the Fermi surface reconstruction through the electron-phonon coupling. 
Thus, the sudden change of \K\ at \TCDW\ is dominated by (i) a sudden reconstruction of the Fermi surface. This is in agreement with the sudden drop of the charge carrier concentrations (Fig.~\ref{fig:MB}(f)).

The mobilities of the individual bands show very strong and non-trivial temperature dependencies Fig.~\ref{fig:MB}(g). At room temperature, the hole mobility is larger than the electron mobility whilst this is reversed at lowest temperatures. 
Both mobilities have a minimum at \TCDW\ naturally associated with strong scattering from the softening mode associated with the CDW formation. Indeed, the temperature dependence of the mobilities show a dip similar in shape to the energy dependence of the L-point phonon mode \cite{Holt2001} (reproduced in Fig.~\ref{fig:MB}(g)).

As noticed by Velebit et al.\ \cite{Velebit2016}, the mobilities of the two bands are roughly equal at \TCDW\ as shown in Fig.~\ref{fig:MB}(h). This equality highlights that scattering from the L-point mode is the dominant process as the phase space for scattering from the electron to the hole band involves the density of states in both bands and the electron-phonon coupling.

The temperature dependence of the separated electron (\sigmae) and hole (\sigmah) conductivity are presented in Fig.~\ref{fig:MB}(i). They show a minimum around $\approx \SI{170}{\kelvin}$ and $\approx\SI{120}{\kelvin}$, respectively. The total conductivity $\sigmat = \sigmae + \sigmah$ and the total resistivity $\rhot = 1/\sigmat$ are dominated by \sigmae\ and \sigmah\ in different temperature regimes. The comparison of the thus calculated \rhot\ and the measured $\rhoxx(H=0)$ in Fig.~\ref{fig:MB}(i) highlights the accuracy of the parameters extracted from the two-band fits. 

From the temperature dependencies of the individual and total conductivities we identify that (i) at high temperature holes dominate the total conductivity and (ii) the negative $\dd\rho /\dd T$ above \TCDW\ is a consequence of the hole mobility increasing with temperature above \TCDW. (iii) The peak in $\rho(T)$ is dominated by the loss of a large portion of charge carrier concentration of both holes and electrons due to the CDW with (iv) the positive $\dd\rho /\dd T$ at low temperatures arising due to the large increase in electron mobility towards lowest temperatures. (v) At low temperatures electrons dominate the conductivity. 
In summary, we conclude that the magnetotransport is dominated by the opening of the CDW gap and the scattering from the underlying bosonic mode.


%
%
Despite the smooth evolution of the resistivity, the magnetotransport behaviour provides direct evidence for an underlying abrupt Fermi surface reconstruction both as a large drop of the charge carrier concentrations and a sudden violation of Kohler's law below \TCDW. The analysis shows the loss of both electrons and holes below \TCDW\ highlighting the strong coupling between them. The fact that a large fraction of the charge carrier concentration is involved in the formation of the CDW enables a clear view at the electronic scattering associated with the L-point CDW mode.  The minimum in the mobility at \TCDW\ is a direct match to the softening of the L-point mode and thus confirms that the dynamics of the charge carriers are directly linked to the dynamics of the CDW mode. Importantly, our measurements show that the strong scattering of the CDW mode causes the intriguing negative $\dd \rho / \dd T$ at room temperature. This makes \TiSe\ uniquely suitable to observe the strong coupling of the CDW mode to the electronic states.
The scattering of electrons from the softening CDW mode is obscured in other prototypical CDW systems like \NbSe\ or \VSe\ where only small portions of the Fermi surface are matched by the ordering wave vector and only these ``hot'' parts experience strong scattering \cite{Rossnagel2001, Strocov2012}. 
 Thus, \TiSe\ is a prototypical system to study the effects of a Fermi surface reconstruction arising from a charge-density wave. These results will be relevant to understand systems like cuprate and iron-pnictide superconductors \cite{Putzke2018,Watson2015}.

%
%
\begin{acknowledgments}
The authors thank Jasper van Wezel, Jans Henke, Antony Carrington, Martin Gradhand, Matthew Wattson, and Phil King for valuable discussion. The authors acknowledge support by the EPSRC under grants EP/N01085X/1, EP/N026691/1, EP/L015544/1, NS/A000060/1, support of the HFML-RU/NWO, member of the European Magnetic Field Laboratory (EMFL), and funding from the European Research Council (ERC) under the European Union’s Horizon 2020 research and innovation programme (Grant agreement No. 715262-HPSuper).
\end{acknowledgments}

The research data supporting this publication can be accessed through the University of Bristol data repository \cite{TiSe2_Data}.

%
%



\section*{Supplementary material}

# Supplementary Information for

# Fermi surface reconstruction and electron dynamics at the charge-density-wave transition in TiSe$_2$

Patrick Knowles,[1] Bo Yang,[1] Takaki Muramatsu,[1] Owen Moulding,[1] Jonathan
Buhot,[2] Charles Sayers,[3] Enrico Da Como,[3] and Sven Friedemann[1, *]

[1] *HH Wills Laboratory, University of Bristol, Bristol, BS8 1TL, UK*

[2] *High Field Magnet Laboratory, University of Radboud, Nijmegen, NL*

[3] *Centre for Nanoscience and Nanotechnology,*

*Department of Physics, University of Bath, BA2 7AY Bath, UK*

(Dated: Friday 20$^{\text{th}}$ March, 2020)

* Sven.Friedemann@bristol.ac.uk

# S I.  SAMPLE PREPARATION

Single crystals were grown by chemical vapour transport (CVT) inside an evacuated quartz ampoule using iodine as the transport agent. The starting materials consisted of high purity titanium ($> 99.9\,\%$) and selenium ($> 99.9\,\%$) powders, and iodine ($> 99.9\,\%$) pellets. Excess selenium was included to encourage the growth of stoichiometric samples [1].

Like all growth methods for $TiSe_2$ employed so far a small level of defects and/or doping influences the electronic properties. Even in the most optimised conditions CVT grown samples show electron doping from iodine substitution on the selenium site and additional titanium as intercalation [1, 2]. Alternative growth methods avoiding iodine show semiconducting behaviour and weak localisation [3–5]. Despite the variation in electronic ground state, the CDW transition has been established for samples grown with different methods. In particular, the transition temperature is identical at $T_{CDW} = 202\,K$ for iodine- and self-flux grown samples with low doping levels as achieved at low growth temperatures [5]. For larger doping levels $T_{CDW}$ is reduced. Indeed, our samples have $T_{CDW} = 202\,K$.

Separate samples were cut with a razor blade with dimensions optimised for measurements of $\rho_{xx}$ and $\rho_{xy}$ from one large single crystal. Figs. S1 and S2 show the basic characterisation of our samples. Contacts have been made with silver paint Dupont 4929N. Samples typically $2\,mm$ long and $0.2\,mm$ wide were used for the measurement of $\rho_{xx}$ and the quantum oscillations in $\rho_{xx}(B)$. Here, all contacts were placed across the width of the sample. Samples typically $1\,mm$ long and $0.5\,mm$ wide were used to measure $\rho_{xy}$. Current contacts were made across the width of the sample while voltage contacts were placed on opposite sides at the middle of the sample. Dimensions were measured with an optical microscope to a precision $\pm 1\,\%$ or $\pm 3\,\mu m$, whichever is larger. Samples were typically $20\,\mu m$ thick. Thus, a systematic error to the charge carrier concentrations of $\pm 15\,\%$ remains. The qualitative results of the 2-band analysis, however, are not affected as only errors of the width and length will effect the ratio of $\rho_{xx}$ and $\rho_{xy}$ which are of the order of $1\,\%$ only.

Samples from 2 batches were used for QO measurements. The data in Fig. 1 of the main manuscript were collected on a sample from a batch grown previously by Wilson et al.[6]. Measurements on samples from the same piece used for the transport studies show an identical quantum oscillation frequency and an identical mean free path was obtained from the Dingle fit (cf. Fig. S3).

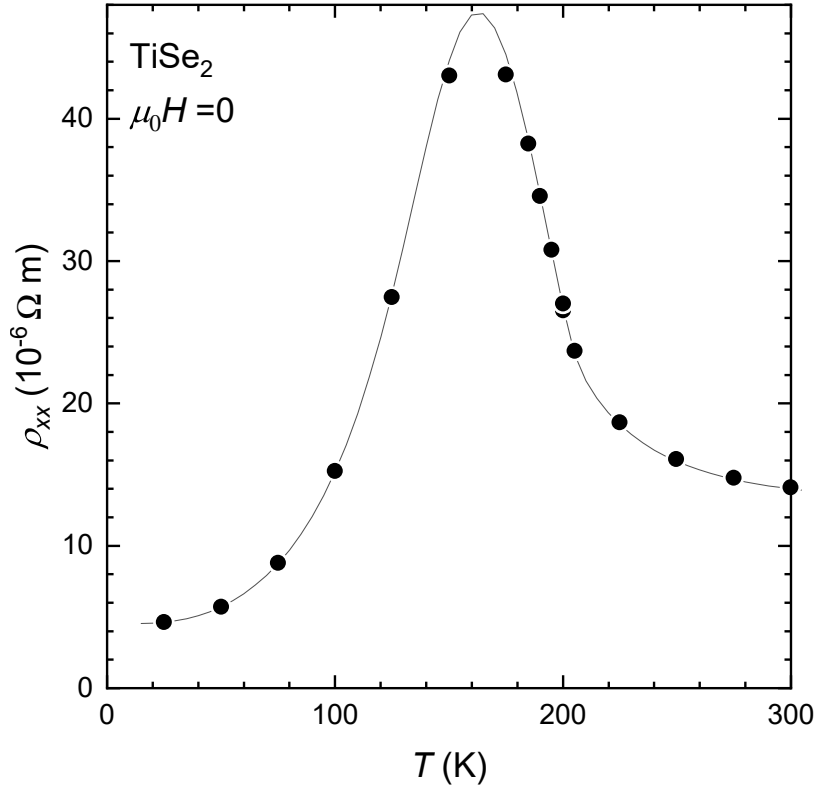

FIG. S1. Resistivity of our TiSe$_2$ samples. Circles correspond to the zero-field values of the sample presented in Fig. 2(b)-(e) of the main manuscript. Solid line corresponds to continuous measurement on a separate sample from the same batch scaled to account for uncertainties in geometry.

Measurements of $\rho_{xx}$ and $\rho_{xy}$ were conducted in a variable temperature insert with a Cernox temperature sensor with negligible magnetoresistance in the temperature range $T \geq 25\,\mathrm{K}$ leading to a relative error of $\rho_{xx}(B)$ less than $3 \times 10^{-5}$. A temperature stability of better than $3\,\mathrm{mK}$ was achieved over the full temperature range.

## S II.   2-BAND ANALYSIS

Motivated by the theoretical band structure and ARPES measurements showing an electron and a hole pocket at high temperatures, we use a two-band model to simultaneously fit the longitudinal and Hall resistivity, $\rho_{xx}(B)$ and $\rho_{xy}(B)$. In order to take into account the fixed doping level (from iodine substitution) of otherwise compensated TiSe$_2$ we rewrite the two-band model in terms of a total charge carrier concentration $n_\mathrm{d} = n_\mathrm{h} - n_\mathrm{e}$ as the balance of hole and electron charge carrier concentrations $n_\mathrm{h}$ and $n_\mathrm{e}$, respectively.

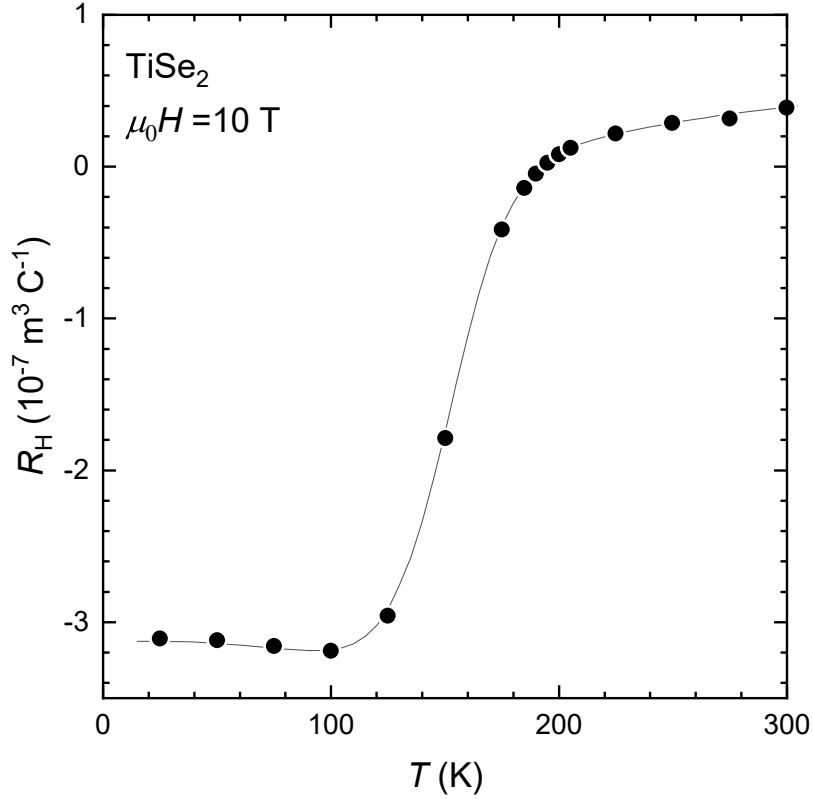

FIG. S2. Hall coefficient our our TiSe₂ samples in a magnetic field of $\mu_0 H = 10\,\mathrm{T}$. Circles correspond to the values of the sample presented in Fig. 2(a) of the main manuscript. Solid line corresponds to continuous measurement on a separate sample from the same batch scaled to account for uncertainties in sample thickness.

$$\rho_{xx} = \frac{1}{e} \frac{\mu_{\mathrm{h}} n_{\mathrm{h}} + \mu_{\mathrm{e}}(n_{\mathrm{h}} - n_{\mathrm{d}}) + \mu_{\mathrm{h}} \mu_{\mathrm{e}} \left[ \mu_{\mathrm{e}} n_{\mathrm{h}} + \mu_{\mathrm{h}}(n_{\mathrm{h}} - n_{\mathrm{d}}) \right] \mu_0^2 H^2}{\left[ \mu_{\mathrm{h}} n_{\mathrm{h}} + \mu_{\mathrm{e}}(n_{\mathrm{h}} - n_{\mathrm{d}}) \right]^2 + \mu_{\mathrm{h}}^2 \mu_{\mathrm{e}}^2 n_{\mathrm{d}}^2 \mu_0^2 H^2} \tag{S1}$$

$$\rho_{xy} = \frac{1}{e} \frac{\left[ \mu_{\mathrm{h}}^2 n_{\mathrm{h}} - \mu_{\mathrm{e}}^2(n_{\mathrm{h}} - n_{\mathrm{d}}) \right] \mu_0 H + \mu_{\mathrm{h}}^2 \mu_{\mathrm{e}}^2 n_{\mathrm{d}} \mu_0^3 H^3}{\left[ \mu_{\mathrm{h}} n_{\mathrm{h}} + \mu_{\mathrm{e}}(n_{\mathrm{h}} - n_{\mathrm{d}}) \right]^2 + \mu_{\mathrm{h}}^2 \mu_{\mathrm{e}}^2 n_{\mathrm{d}}^2 \mu_0^2 H^2} \tag{S2}$$

The mobilities of the hole and electron band are given by $\mu_{\mathrm{h}}$ and $\mu_{\mathrm{e}}$, respectively, $e$ denotes the absolute value of the electron charge, $\mu_0$ the vacuum permeability, and $H$ the magnetic field.

Starting values for the fitting of the two-band model were identified from separated hole and electron conductivities at lowest temperature using a Kramers-Kronig transformation as employed for the mobility spectral density analysis [7]. We determined $n_{\mathrm{d}} \approx 1.5 \times 10^{25}\,\mathrm{m}^{-3}$ from the fits at lowest temperatures $T \leq 50\,\mathrm{K}$ and fix this value for all temperatures.

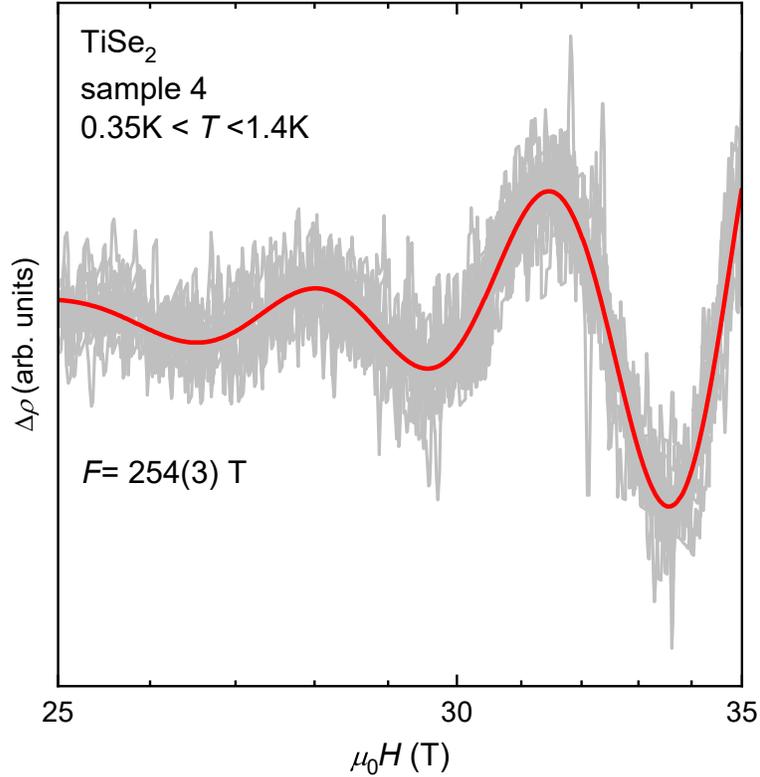

FIG. S3. Quantum oscillations in sample 3 of TiSe$_2$. Data at several temperatures below 0.4 K have been overlaid. No significant temperature dependence is observed in this temperature range. The solid red line is a fit with a single frequency and exponential damping and including a 3rd-order polynomial background.

## S III. ANALYSIS OF ANGULAR DEPENDENT QUANTUM OSCILLATION FREQUENCY

The angular dependence of the quantum oscillation frequency $F(\theta)$ shown in Fig. 1(d) of the main manuscript has been fitted to a model ellipsoid Fermi surface shape with an in-plane semi-axis $k_{ab}$ and an out-of-plane semi-axis $k_c$ using

$$F(\theta) = F_{ab} \times \frac{r}{\sqrt{r^2 \cos^2 \theta + \sin^2 \theta}} \tag{S3}$$

where the in frequency for in-plane orbits is given by

$$F_{ab} = \frac{\hbar A_{ab}}{2\pi e}$$

with an in-plane cross section $A_{ab} = \pi k_{ab}^2$ and the ratio of the semi-axes $r = k_c/k_{ab}$.

| | $k_{ab}$ | $r$ | $m_e^\star$ | $l$ | $\gamma$ | Ref. |
|---|---|---|---|---|---|---|
| | ($\text{Å}^{-1}$) | (-) | ($m_e$) | (nm) | ($\text{mJ}\,\text{mol}^{-1}\,\text{K}^{-2}$) | |
| QO | 0.09 | 1.5 | 0.62(4) | 9(1) | 0.18 | |
| 2-band transport | 0.07 | | | 40(2) | | |
| ARPES | 0.08 | 1.6 | 0.9(2) | | | [9] |
| heat capacity | | | | | 0.18 | [10] |

TABLE S1. Low-temperature Fermi surface characteristics of the electron pocket in $TiSe_2$. The in-plane semi-axis of the Fermi surface is denoted as $k_{ab}$ the ratio of the two semi-axes as $r$. The effective mass and mean free path are given by $m_e^\star$ and $l$, respectively. The Sommerfeld coefficient $\gamma$ was calculated in a free-electron picture taking into account the size of the electron pocket and the effective mass. The transport parameters have been calculated from $n_e(T = 25\,\text{K})$ and $\mu_e(T = 25\,\text{K})$ extracted from the 2-band fits assuming $r = 1.5$ (as obtained from the QO analysis). Errors are standard errors of the Lifshitz–Kosevich and Dingle fits to the QO data. The error of the transport mean free path includes uncertainties in the dimensions of the sample.

The ellipsoid form used to fit the angular dependence of $F(\theta)$ is consistent with the 6-fold symmetry dictated by the low-temperature reconstruction of the Brilloiun zone where the L-point is folded back to the $\Gamma$-point [8, 9].

The parameters of the electron pocket extracted from the quantum oscillation measurements are compared to our transport measurements, ARPES [9] and heat capacity [10] studies in table S1.

---



\end{document}